\documentclass[12pt,fleqn]{article}
\usepackage{amssymb}
\usepackage{hyperref}
\usepackage{amsmath}

\usepackage{amsmath,amssymb,amsthm}

\begin{document}
\begin{center}
{\Large \textbf{Eikonal 2.0: General Solutions for Modified Eikonal Equations}}

\vskip 20pt {\large \textbf{Iryna YEHORCHENKO}}

\vskip 20pt {Institute of Mathematics of NAS Ukraine, 3 Tereshchenkivs'ka Str., Kyiv-4, Ukraine} \\
E-mail: iyegorch@imath.kiev.ua
\\
Institute of Mathematics, PAN, Poland\\
iyehorchenko@impan.pl
\end{center}

\footnotesize
\begin{abstract}
We find implicit general solutions for modified eikonal equations
$u_a u_a=F(u_t)$,
where lower indices at  dependent variables designate derivatives, $a=1,2,..,n$, and summation is implied over the repeated indices.

We will consider general solutions and symmetries of these equations, and relations of these equations to reduction procedures of higher-order PDE.

\end{abstract}

\normalsize
\section{Introduction}
This paper is a prolongation of the paper \cite{preprintIY2022a}, one in the series intended as comprehensive study of the eikonal equations and systems that include them.

We will consider here a class of nonlinear first-order partial differential equations (PDE) in the following form:
\begin{equation}\label{eik1class}
u_a u_a =F(u_t),
\end{equation}
\noindent
where in general we have one real-valued dependent variable $u=u(x_0,x_1,\ldots,x_n)$, a time variable $x_0$ and $n$ space variables $x_1,\ldots,x_n$. Thus class includes both the standard eikonal equation and the Hamilton-Jacobi equation. Here we do not consider cases when $F(u_t)=\rm{const}$.

I would like to point out that $u$ and other dependent variable take values in a real space, as the complex-valued dependent variables require different treatment and a special study.

The intended audience of this paper is wider than "the symmetry analysis people", so many detailed arrangements and calculations that are obvious for the above group are written explicitly. The results in this paper I believe at the moment of writing to be new will be adduced at the end. I mentioned only the literature what was immediately relevant for the paper. Further we will mostly adhere to notations introduced in \cite{preprintIY2022a}.

Group classification of the class 
\begin{equation}\label{eik2class}
u_a u_a =F(t,u,u_t),
\end{equation}
the class (\ref{eik1class}) we consider being its subclass, was performed in \cite{preprintIY-RP2001}. We adduce here this classification.

\begin{tabular}{|r|p{3.4cm}|p{10cm}|}
\hline
&\hfil$F$\hfil&\hfil Basis Symmetry Operators \hfil \\ \hline
$\!0\!$&$F(t,u,u_t)$&$\partial_a$, $J_{ab}$\\ \hline
$\!1\!$&$e^{\delta t}\!f(u,u_t)$, $\!\delta\!\in\!\{0;1\}\!\!\!$&$\partial_a$, $J_{ab}$, $2\partial_t-\delta x_a\partial_a$\\ \hline
$\!2\!$&$e^u h(u_t)$&$\partial_a$, $J_{ab}$, $\partial_t$, $2\partial_u-x_a\partial_a$\\ \hline
$\!3\!$&$|u|^{2-\delta} h(u_t)$, $\delta\not=2$&$\partial_a$, $J_{ab}$, $\partial_t$, $2t\partial_t+2u\partial_u+\delta x_a\partial_a$\\ \hline
$\!4\!$&$h(u_t)$&$\partial_a$, $J_{ab}$, $\partial_t$, $\partial_u$, $D$\\ \hline
$\!5\!$&$e^{u_t}$&$\partial_a$, $J_{ab}$, $\partial_t$, $\partial_u$, $D$, $x_a\partial_a-2t\partial_u$\\ \hline
$\!6\!$&$|u_t|^\beta$, $\beta\not=0,1,2$&$\partial_a$, $J_{ab}$, $\partial_t$, $\partial_u$, $D$,
    $(\beta-2)x_a\partial_a-2u\partial_u$\\ \hline
$\!7\!$&$u_t^2$&
$2g_{\mu\nu}c^\mu(u)x_\nu x_\varkappa\partial_\varkappa
-c^\varkappa(u)g_{\mu\nu}x_\mu x_\nu\partial_\varkappa
+g_{\mu\nu}b^{\mu\varkappa}(u)x_\nu\partial_\varkappa
+$\hfil${ }$
$+d(u)x_\varkappa\partial_\varkappa+a^\varkappa(u)
\partial_\varkappa+\eta(u)\partial_u$ \\ 
$\!8\!$&$\varepsilon_2 u_t^2+\varepsilon_1$&
$\partial_a$, $J_{ab}$, $\partial_t$, $\partial_u$, $D$,
$J_{ua}=u\partial_a+\varepsilon_1x_a\partial_u$, $J_{ta}=t\partial_a+\varepsilon_2x_a\partial_t$,
$J_{ut}=u\partial_t-\varepsilon_1\varepsilon_2t\partial_u$,
$K_a=2x_aD-s^2\partial_a$, $K_u=2uD+\varepsilon_1s^2\partial_u$,
$K_t=2tD+\varepsilon_2s^2\partial_t$,
 where $s^2=x_ax_a-\varepsilon_1u^2-\varepsilon_2t^2$\\ \hline
$\!9\!$&$\varepsilon_2 e^uu_t^2+\varepsilon_1$&
$\partial_a$, $J_{ab}$, $\partial_t$, $t\partial_t+2\partial_u$,
$(t^2-4\varepsilon_1\varepsilon_2e^u)\partial_t+4t\partial_u$ \\ \hline
$\!10\!$&$\cos^{-2}\!u\,u_t^2+1$&
$\partial_a$, $J_{ab}$, $\partial_t$, $\cos t\tan u\,\partial_t-\sin t\,\partial_u$, $\sin t\tan u\,\partial_t+\cos t\,\partial_u$ \\ \hline
$\!11\!$&$\pm(\cos^{-2}\!u\,u_t^2-1)$&
$\partial_a$, $J_{ab}$, $\partial_t$, $\cosh t\tan u\,\partial_t+\sinh t\,\partial_u$, $\sinh t\tan u\,\partial_t+\cosh t\,\partial_u$ \\ \hline
$\!12\!$&$\cosh^{-2}\!u\,u_t^2+1$&
$\partial_a$, $J_{ab}$, $\partial_t$, $\cosh t\tanh u\,\partial_t-\sinh t\,\partial_u$, $\sinh t\tanh u\,\partial_t-\cosh t\,\partial_u$ \\ \hline
\end{tabular}
\vskip20pt
$F=F(t,u,u_t)$, $f=f(u,u_t)$, $h=h(u_t)$ are arbitrary smooth functions of their arguments,
$\delta$ is constant, $\varepsilon_1,\varepsilon_2=\pm 1$,
$(\varepsilon_1,\varepsilon_2)\not=(-1,-1)$,
$J_{ab}=x_a\partial_b-x_b\partial_a$, $D=t\partial_t+u\partial_u+x_a\partial_a$.
In the case~7, $g_{\mu\nu}$ is metric tensor of Minkowsky space $\mathbb{R}^{1,n}$,
i.e. $g_{00}=-g_{11}=-g_{22}=-g_{33}=1$, $g_{\mu\nu}=0$, $\mu\not=\nu$;
$c^\mu$, $b^{\mu\varkappa}$, $d$, $a^\varkappa$, $\eta$ are arbitrary smooth functions of the
variable $u$ (for operators of such form to form an algebra,
it is necessary to require that these functions were infinitely differentiable or real analytical);
the indices $\mu$, $\nu$ and $\varkappa$ take values from 0 to 3.

Let us note that the known Hamilton-Jacobi and eikonal equations, and the relativistic Hamilton
equation are selected as representatives of the classes of equivalent equations having the widest symmetry.

\section{General Solutions}
In consideration of the general solutions for the eikonal equations, we will follow the idea, given in \cite{FZhR-preprint}, where also general solutions for the standard eikonal equations
$ u_\mu u_\mu =1$ and  $ u_\mu u_\mu =0$ 
for $n=3$ are adduced. 

Here we  generalize the relevant formulae the modified eikonal equations and for the arbitrary number of space variables.

We used ideas for finding general solutions outlined e.g. in \cite{FZhR-preprint}. The main tool of this approach is to use contact transformations that are  nonlocal transformations involving dependent and independent variables, and also first derivatives of dependent variables. 

We will apply the following format of such contact transformations, the indices at independent variables $x'_{\mu}$, $x_{\mu}$ are their indices indicating their respective numbers, and the indices at dependent variables $u'_{\mu}$, $u_{\mu}$ designate their derivatives with respect to the relevant independent variables $x'_{\mu}$ or $x_{\mu}$.
\begin{gather} \nonumber
  x'_{\mu}= \varphi_{\mu}(x_{\mu},u,u_{\mu},)\\ \label{contactgen}
  u' = \Phi(x_{\mu},u,u_{\mu}),\\ \nonumber
  u'_{\mu} =\Phi_{\mu}(x_{\mu},u,u_{\mu}).
\end{gather}

Here $x'_{\mu}$, $u'$ are new transformed variables,
 $u'_{\mu}$ are transformed first derivatives, $\varphi_{\mu}$, $\Phi$ and $\Phi_{\mu}$ are sufficiently smooth functions on their variables. Note that indices at $\Phi_{\mu}$ are just numerating indices, not derivatives.

We can apply the partial derivative operators to (\ref{eik1class}) and conclude linear dependence of the resulting equations considered as linear equations for the first derivatives of $u$, and that the rank of the matrix of the second derivatives of the function $u$ may take values from 0 to $n$,  where $n$ is the number of space variables.

The equation $u_a u_a =F(u_t)$ has real solutions for which the rank of the matrix of the second derivatives is zero, all first derivatives $u_{\mu}$ are constants, and $u$ is linearly dependent on $x_{\mu}$:
\begin{equation} \nonumber
 u=c_{\mu}x_{\mu}+c,
\end{equation}
\noindent
where $c_{\mu}$ and $c$ are arbitrary constants such that $c_{a}c_{a}=F(c_0)$, when $F(c_0)$ is non-negative.

In the case when the rank of $U$ is equal to $n$, we can apply the contact transformations
\begin{gather} \nonumber
  H=x_a u_{x_a}-u, \qquad y_a=u_{x_a},
  \qquad t=y_0;\\  \nonumber
  H_{y_a}=x_a, \qquad
  H_{y_0}=-u_t; \nonumber
 \end{gather}
\noindent
and obtain the following equation for the new function $H$:
\begin{equation} \nonumber
y_a y_a = F(H_{y_0}).
\end{equation}
We can linearize it with respect to $H_{y_0}$: $H_{y_0}=\Phi(y_b y_b)$,
where $\Phi$ is a inverse function to $F$. Its solution is
$$ H=y_0\Phi(y_a y_a)-\Psi(y_d),$$
\noindent
where $\Psi(y_d)$ is a sufficiently smooth arbitrary function. Further we use parameters $\tau_d$ instead of $y_d$.
An $n$-parameter general solution (${\rm rank}(U) =n$):
\begin{gather} \label{gseik1-nn}
  u=-x_a\tau_a+t\Phi(\tau_b \tau_b)+\Psi(\tau_d), \\ \nonumber
   x_a-t\Phi_{\tau_a}
   (\tau_b \tau_b)-\Psi_{\tau_a}(\tau_d) =0 \nonumber
 \end{gather}
\noindent
where $\Psi(\tau_d)$ is a sufficiently smooth arbitrary function on $n$ parameters $\tau_d$.

The formula (\ref{gseik1-nn}) includes arbitrary functions, so all solutions (\ref{gseik1-nn}) cannot be equivalent to symmetry solutions.
\vskip10pt
In general (not always), formulas for the general solutions give wider classes of solutions than the Lie symmetry method.
As to the zero-rank general solution, all such solutions are symmetry solutions.

General solutions depending on $k$ parameters may be taken as follows:
\begin{gather} \nonumber
  u=-x_b\tau_b+t\Phi(\tau_d\tau_d+w_mw_m)+w_mx_{k+m}
  +\Psi(\tau_d), \\ \nonumber
   x_b-t
   \Phi_{\tau_b}(\tau_d\tau_d+w_mw_m)
   -w_{m\tau_b}x_{k+m}-\Psi_{\tau_b}(\tau_d) =0,
 \end{gather}
\noindent
where $\Psi(\tau_d)$ is a sufficiently smooth arbitrary function on $n$ parameters $\tau_d$.

\section{Examples for Specific Equations}
We will write an example using one of the equations from the group classification table:
\begin{equation} \nonumber
u_a u_a =\exp(u_t),
\end{equation}
\noindent
or
\begin{equation}\label{ex2}
u_t=\ln(u_a u_a).
\end{equation}

Equation (\ref{ex2}) has real solutions for which the rank of the matrix of the second derivatives is zero, all first derivatives $u_{a}$ and $u_t$ are constants, and $u$ is linearly dependent on $x_{\mu}$:
\begin{equation} \nonumber
 u=c_{\mu}x_{\mu}+c,
\end{equation}
\noindent
where $c_{\mu}$ and $c$ are arbitrary constants such that $c_{a}c_{a}=\exp(c_0)$.

An $n$-parameter general solution (${\rm rank}(U) =n$):
\begin{gather} \nonumber
  u=-x_a\tau_a+t\ln(\tau_b \tau_b)+\Psi(\tau_d), \\ \nonumber
   x_a-t\frac{2\tau_a}{\tau_b \tau_b}
   -\Psi_{\tau_a}(\tau_d) =0, \nonumber
 \end{gather}
\noindent
where $\Psi(\tau_d)$ is a sufficiently smooth arbitrary function on $n$ parameters $\tau_d$.

General solutions depending on $k$ parameters may be taken as follows:
\begin{gather} \nonumber
  u=-x_b\tau_b+t\ln(\tau_d\tau_d+w_mw_m)+w_mx_{k+m}
  +\Psi(\tau_d), \\ \nonumber
   x_b-t
   \frac{2\tau_b}{\tau_c \tau_c}
   -w_{m\tau_b}x_{k+m}-\Psi_{\tau_b}(\tau_d) =0,
 \end{gather}
\noindent
where $\Psi(\tau_d)$ is a sufficiently smooth arbitrary function on $n$ parameters $\tau_d$.

\section{New Results, Conclusions and Further Work}
The new results of this paper are general solutions of modified eikonal equations for arbitrary number of space variables.

There are multiple papers searching for numerical solutions of the boundary problems for eikonal and similar first-order partial differential equations. On my opinion, it would be interesting to look for these using the general solutions.

Further work will include study of the coupled modified eikonal equations in higher dimensions, relevant conditional and hidden symmetries, and looking for higher-order equations with the same or similar symmetry properties as modified eikonal equations.

\section{Acknowledgements}
The first and foremost my acknowledgements go to the
Armed Forces of Ukraine and to the Territorial Defence Forces of Ukrainian Regions due to whom I am alive and is able to work.

Please remember that Russia is an aggressor country and still plans to kill all Ukrainians not going to be their subordinates, and Russian scientists contribute to killings despite words about peace from a tiny portion of them; if even by trying to persuade na\"{i}ve people that Russia is a civilised country.

This research was mostly completed during my work at the Institute of Mathematics of the Polish Academy of Sciences with the grant of the Narodowe Centrum Nauki (Poland), Grant No.2017/26/A/ST1/00189. I would like to thank the Institute of Mathematics of the Polish Academy of Sciences for their hospitality and grant support, to the National Academy of Sciences of the USA and the National Centre of Science of Poland for their grant support.

Further work was supported by a grant from the Simons Foundation (1290607, I.A.Y).

\end{document}